\documentclass[a4paper,11pt]{article}

\usepackage[top=1.0in,right=0.8in,left=0.8in,bottom=1.75in]{geometry}

\usepackage{amsmath,amsfonts,amssymb}
\usepackage{booktabs}
\usepackage{caption}
\usepackage{enumerate}
\usepackage{graphics,graphicx}
\usepackage{latexsym}
\usepackage{multirow}
\usepackage{pst-all}
\usepackage{sectsty}
\usepackage{subfig}
\usepackage{tabularx}
\usepackage[affil-it,auth-sc]{authblk}
\allsectionsfont{\normalsize}

\newcommand{\beq}{\begin{equation}}
\newcommand{\eeq}{\end{equation}}
\newcommand{\lb}{\label}
\newcommand{\beqar}{\begin{eqnarray}}
\newcommand{\eeqar}{\end{eqnarray}}
\newcommand{\bit}{\begin{itemize}}
\newcommand{\eit}{\end{itemize}}
\newcommand{\benum}{\begin{enumerate}}
\newcommand{\eenum}{\end{enumerate}}
\newcommand{\barr}{\begin{array}}
\newcommand{\earr}{\end{array}}
\def\ds{\displaystyle}

\newenvironment{sistema}{\left \lbrace \begin{array}{@{}l@{}}}{\end{array} \right.}

\def\diff{d}

\def\XXint#1#2#3{{\setbox0=\hbox{$#1{#2#3}{\int}$}
   \vcenter{\hbox{$#2#3$}}\kern-.5\wd0}}

\def\b0{\mbox{\boldmath $0$}}

\def\bm{\mbox{\boldmath $m$}}
\def\bn{\mbox{\boldmath $n$}}

\def\bt{\mbox{\boldmath $t$}}

\def\bI{\mbox{\boldmath $I$}}

\def\bS{\mbox{\boldmath $S$}}

\newcommand{\bsigma}{\mbox{\boldmath $\sigma$}}

\newcommand{\bepsilon}{\mbox{\boldmath $\epsilon$}}

\def\f0{\ensuremath{\mathbb{O}}}

\newcommand{\tr}{\mathop{\mathrm{tr}}}

\def\AMEC{{\it Acta Mech.}\ }

\def\CMAME{{\it Comput.\ Method.\ Appl.\ M.}\ }
\def\EFM{{\it Eng.\ Fract.\ Mech.}\ }
\def\EJMA{{\it Eur.\ J.\ Mech. A-Solid.}\ }

\def\IJF{{\it Int.\ J.\ Fracture}\ }
\def\IJMS{{\it Int.\ J.\ Mech.\ Sci.}\ }

\def\IJP{{\it Int.\ J.\ Plasticity}\ }
\def\IJSS{{\it Int.\ J.\ Solids Struct.}\ }

\def\JMPS{{\it J.\ Mech.\ Phys.\ Solids}\ }

\def\MMS{{\it Math.\ Mech.\ Solids}\ }
\def\MOM{{\it Mech.\ Materials}\ }

\title{Integration algorithms of elastoplasticity \\ 
for ceramic powder compaction}

\author[]{M. Penasa}
\author[]{A. Piccolroaz}
\author[]{L. Argani}
\author[]{D. Bigoni\footnote{Corresponding author:\,e-mail:\,bigoni@ing.unitn.it; phone:\,+39\,0461\,282507.}}
\affil[]{Department of Civil, Environmental and Mechanical Engineering \\ University of Trento \\ Via Mesiano 77, 38123 Trento, Italy}

\date{}

\begin{document}

\maketitle

\begin{abstract}

Inelastic deformation of ceramic powders (and of a broad class of rock-like and granular materials), can be described with the yield function proposed by Bigoni and Piccolroaz (2004,  Yield criteria for quasibrittle and frictional materials. Int. J. Solids and Structures, 41, 2855-2878). This yield function 
is not defined outside the yield locus, so that \lq gradient-based' integration algorithms of elastoplasticity cannot be directly 
employed. Therefore, we propose two {\it ad hoc} algorithms: (i.) an explicit integration scheme based on a forward Euler technique with a 
\lq centre-of-mass' return correction and (ii.) an implicit integration scheme based on a \lq cutoff-substepping' return algorithm. 
Iso-error maps and comparisons of the results provided by the two algorithms with two exact solutions (the compaction 
of a ceramic powder against a rigid spherical cup and the expansion of a thick spherical shell made up of a green body), show that both the proposed algorithms perform correctly and accurately.

\end{abstract}


\section{Introduction}

Granular and geological materials
are employed for many industrial purposes: shock and vibration absorbers, fire protection, thermal barriers, refractory products, wear protectors, electric isolators, and catalysts. They are 
characterized by pressure-sensitive yielding and dilatant/contractant inelastic behaviour\footnote{
These mechanical behaviours are observed in: 
ceramic and metal powders \cite{pibiga1,pibiga2,bier,ha,hei}, 
concrete \cite{ba}, 
geomaterials \cite{conti,paluz,dal1,dal,dalpino,des,dr,mai,mort,she,zho}, 
masonry \cite{ang,defaveri,wang}, 
but also metals \cite{bo,copp1,copp,ebno,hu,miro,pie,wier}, 
high strength alloys \cite{bai}, 
and shape memory alloys \cite{lex,le,rani98,lex2002,rani,saint,sed,ta}.
}.
Several yield functions have been introduced for the mechanical description of these materials, which have to satisfy different requirements, among which, the 
most important are convexity and smoothness, two requisites met by the yield function proposed by Bigoni and Piccolroaz \cite{bigp} (see also \cite{pibi,bigoni}), henceforth referred to as the \lq BP yield function'. 
Moreover, this function has an extreme \lq deformability', thus results particularly appropriate to describe the granular/solid transition occurring during 
forming of ceramic powders \cite{bosi,stupk}, a crucial process in the production of many ceramic products. 

Used in the context of elastoplastic modelling, the BP yield function introduces the problem that 
to be convex, it has been defined $+\infty$ outside certain regions in the stress-space.
Therefore, in its original form, the PB yield function cannot be implemented within an elastoplastic integration scheme, if a gradient-based return-mapping 
algorithm is used, for which the gradient of the yield function is needed everywhere in the stress-space \cite{sihug}. 
If a non-convex version of the BP yield function (obtained by squaring the terms) is implemented with a return-mapping algorithm, wrong results can be produced, 
as a specific example will demonstrate. 

The problem of the BP yield surface is also common to other yield surfaces for geomaterials \cite{bran}, so that the aim of the present article is to overcome the difficulty by proposing two algorithms: one is based on a forward Euler technique with a correction based on a \lq 
centre-of-mass' return scheme, fully applicable to the original form of the BP yield function (defined $+\infty$ outside the yield surface), and another based on a cutoff-
substepping return-mapping algorithm that can be applied on the squared (and non-convex) version of the BP yield function. 
Iso-error maps and comparisons with two model problems allowing for a semi-analytical solution (the forming of a ceramic powder pressed against a rigid spherical 
cup  and the expansion of a green body spherical shell subject to internal pressure) show that both algorithms perform correctly, with an accuracy comparable in 
certain regions of the stress state, even if there are regions where each algorithm is superior to the other. In particular, the \lq centre-of-mass' algorithm is 
faster that the other, but less accurate near vertices of the deviatoric yield surface, while the cutoff-substepping return-mapping algorithm is always more 
accurate than the other, but can become slow for stress states near the vertices of the meridian yield surface.
Finally, we may conclude that, although both algorithms have their advantages and limitations, generally speaking the cutoff-substepping return-mapping algorithm 
can eventually be preferred.

\section{The \lq centre of mass' integration algorithm}

\begin{figure}[!htcb]
\begin{center}
\includegraphics[width= 100mm]{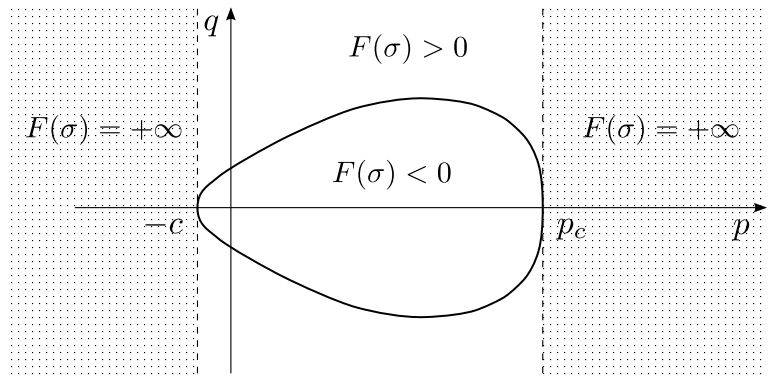}
\caption{\footnotesize{The BP yield function represented as a surface in the $p$--$q$ plane.}}
\label{infinito3}
\end{center}
\end{figure}

As mentioned in the introduction, the problem with the BP yield function is that it is defined $+\infty$ in some regions outside the elastic domain (for 
$p\notin [-c, p_c]$), Fig. \ref{infinito3}. Therefore, an integration algorithm based on a standard return mapping technique cannot work, so that the purpose of this Section is to introduce an explicit forward Euler algorithm to solve
this problem (while an implicit algorithm will be presented in the next Section defined on  a \lq squared version' of the yield function).
%

\subsection{The BP yield surface and its centre of mass}

Bigoni and Piccolroaz \cite{bigp} (see also \cite{pibi}) have introduced a new yield function for \textit{isotropic} materials (called \lq BP' in 
the following), defined in terms of the stress tensor $\bsigma$ by
\begin{equation}
\lb{funzionazza}	
F(\bsigma) = f(p) + \frac{q}{g(\theta)},
\end{equation}
where, defining the parameter $\Phi$ as 
\begin{equation}
\lb{fiegi}
\Phi = \frac{p+c}{p_{c}+c},
\end{equation}
the meridian and deviatoric functions are respectively written as\footnote{
The expression \eqref{effedip}$_2$ of $g(\theta)$ was proposed by Podg\'{o}rski \cite{podo1,podo2} and independently by Bigoni and Piccolroaz \cite{bigp}.
}
\begin{equation}
\lb{effedip}
f(p) =
\left\{
\barr{ll}
-Mp_c\sqrt{\left(\Phi-\Phi^m\right) \left[2(1-\alpha)\Phi+\alpha\right]}, & \Phi \in [0,1], \\[3mm]
+\infty, & \Phi \notin [0,1],
\earr
\right.
\quad
\frac{1}{g(\theta)} = \cos \left[ \beta\frac{\pi}{6} - \frac{\cos^{-1} \left( \gamma \cos 3\theta \right)}{3} \right],
\end{equation}
in which $p$, $q$ and $\theta$ (the Lode's angle) are the following stress invariants
\begin{equation}
\lb{lodazzo}
p = -\frac{\tr \bsigma}{3}, \quad q = \sqrt{3 J_{2}}, \quad \theta = \frac{1}{3} \arccos \left( \frac{3\sqrt{3}}{2} \frac{J_{3}}{J_{2}^{3/2}} \right),
\end{equation}
functions of the second and third invariant of the deviatoric stress $\bS$
\begin{equation}
J_{2} = \frac{1}{2} \tr \bS^{2} ,  \quad  J_{3} = \frac{1}{3} \tr \bS^3 , \quad  \bS = \bsigma - \frac{\tr\bsigma}{3} \bI,
\end{equation}
$\bI$ being the identity tensor.

The yield function \eqref{funzionazza}--\eqref{effedip} is convex when the seven material parameters defining the meridian shape function $f(p)$ and the 
deviatoric shape function $g(\theta)$ lie within the following intervals
\begin{equation}
M > 0, \quad p_c > 0, \quad c \geq 0, \quad 0 < \alpha < 2, \quad m > 1, \quad 0 \leq \beta \leq 2, \quad 0 \leq \gamma \leq 1.
\end{equation}

\paragraph{Centre of mass of the yield surface}\

\begin{figure}[!htcb]
\centering
\includegraphics[width=55mm]{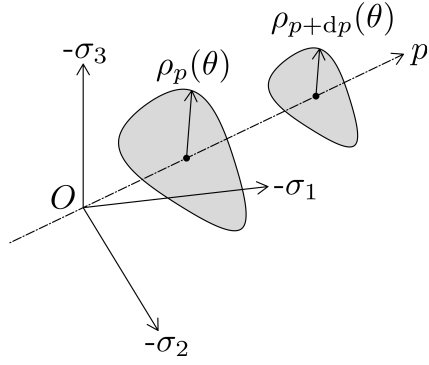}
\caption{Radius $\rho(\theta)$ and centre of mass of two indicative deviatoric sections (located at different mean stresses $p$) of the BP yield surface. 
Due to the isotropy symmetries of the deviatoric sections, the mass centres lie on the hydrostatic axis.}
\label{Masse}
\end{figure}

The numerical integration algorithm that will be developed later is based on the knowledge of the centre of mass of the yield surface. This, with reference 
to Fig.\ \ref{Masse}, can be obtained as follows.
%

We begin noting that the yield surface possesses the isotropy  symmetries in the deviatoric plane (see \cite{bigp}), therefore, the centre of 
mass of the yield surface lies on the hydrostatic axis. The infinitesimal area of the deviatoric section can be evaluated as
\begin{equation}
dA = \frac{1}{2} \rho^{2} (\theta) \, d \theta ,
\end{equation}
where
\begin{equation}
\rho (\theta) = \sqrt{\frac{2}{3}} q = - \sqrt{\frac{2}{3}} f(p) g(\theta),
\end{equation}
is the radius of the surface boundary evaluated with respect to the hydrostatic axis, so that the area of the deviatoric section is expressed as
\begin{equation}
A(p) = 2 f^{2} (p) \! \int_{0}^{\frac{\pi}{3}} \! g^{2}(\theta) \, d \theta.
\end{equation}
On application of the definition of the centre of mass 
\begin{equation}
p_{G} = \frac{\ds \int_{-c}^{p_{c}} p \, A(p) \, dp }{\ds \int_{-c}^{p_{c}} A(p) \, dp },
\end{equation}
provides the \textit{coordinate of the centre of mass of the BP yield surface along the hydrostatic axis}
\begin{equation}
\label{baricentro}
p_{G} =  \frac{ (m+1) p_{c} \left[ (\alpha-3)m-6 \right] + c \left[ 6(\alpha+1)+m(m+7) \right] }{ (m+3) \left[ (\alpha-4)m -2(\alpha+2) \right]},
\end{equation}
a formula involving all the \textit{meridian} parameters of the yield function, except $M$.

\subsection{The \lq centre of mass' return algorithm}

\begin{figure}[!htb]
\centering
\includegraphics[width=65mm]{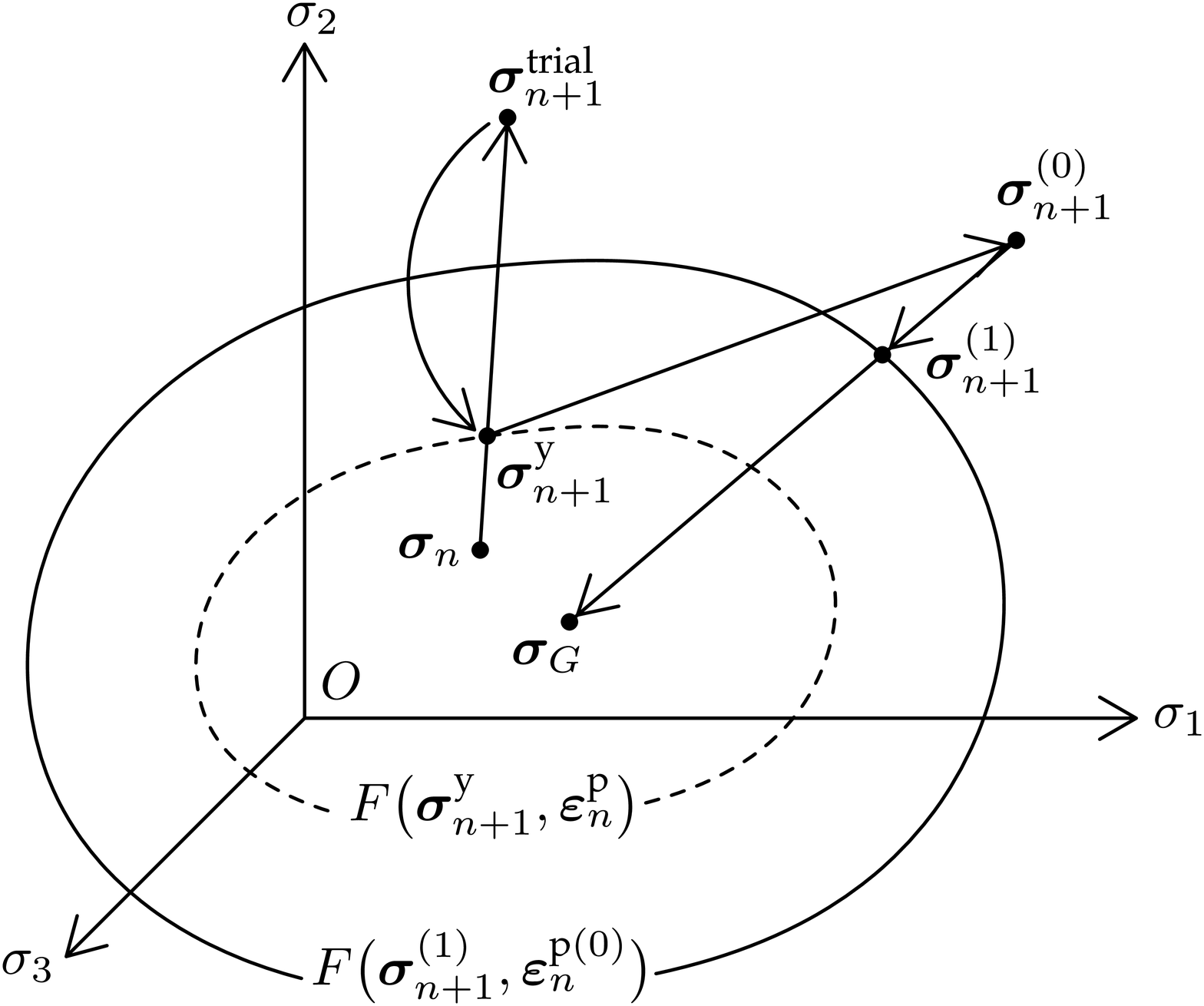}
\caption{Geometrical sketch of the \lq centre of mass return algorithm' for the integration of rate elastoplastic constitutive equations.}
\label{return}
\end{figure}

We propose a numerical integration procedure for rate elastoplastic constitutive equations based on a return algorithm which is geometrically sketched in 
Fig.\ \ref{return} and can be syntetically described with reference to Box 1.
%
%
In particular, starting from a given state at a step $n$ [point (1) in Box 1] and after the usual trial elastic step [point (2)], the stress point at 
yielding is found along the line joining the trial and the initial state [point (3)]; from this point, after the purely elastic strain is eliminated from 
the strain increment [point (4)], a new stress increment is found using the tangent elastoplastic operator [point (5)]; the plastic strain increment is 
updated [point (6)]; and finally, a return on the updated yield surface is performed along the line joining with the centre of mass of the yield surface 
[points (7)--(8)].

\begin{center}
\framebox{
\begin{minipage}{0.8\columnwidth}

\paragraph{\hfill Box 1: The \lq centre of mass' integration algorithm \hfill}

\begin{enumerate}[(1)]

\item Given an initial state at step $n$, described by the variables  $\bsigma_{n}$, $\bepsilon^{e}_{n}$, $\bepsilon^{p}_{n}$ and given a strain 
increment $\Delta \bepsilon$;

\item evaluate the elastic trial solution 
$$ 
\bsigma^{\textup{trial}}_{n+1} = \bsigma_{n} + \mathbb{E} [\Delta \bepsilon];
$$

\item along the line from $\bsigma_{n}$ to $\bsigma^{\textup{trial}}_{n+1}$ find the stress point $\bsigma^{y}_{n+1}$ at yielding \label{puta}
$$
F \left( \bsigma^y_{n+1}, \bepsilon^p_n \right) = 0;
$$ 

\item evaluate the elastic deformation increment corresponding to $\bsigma^{y}_{n+1}-\bsigma_{n}$
$$
\Delta\bepsilon^{y}_{n+1} = \mathbb{E}^{-1}\left[ \bsigma^{y}_{n+1}-\bsigma_{n} \right];
$$

\item evaluate the stress increment via the tangent elastoplastic operator \label{ueppla}
$$ 
\bsigma^{(0)}_{n+1} = \bsigma^{y}_{n+1} + \mathbb{C} \left[ \Delta \bepsilon-\Delta \bepsilon^{y}_{n+1} \right];
$$

\item update the plastic deformation
$$ 
\bepsilon^{p~(0)}_{n+1} = \bepsilon^{p}_{n} + \Delta \bepsilon- \mathbb{E}^{-1} \bigl[ \bsigma^{(0)}_{n+1} - \bsigma_{n} \bigr];
$$

\item find the stress $\bsigma^{(1)}_{n+1}$ on the updated yield surface \label{zorra}
$$ 
F \bigl( \bsigma^{(1)}_{n+1}, \bepsilon^{p~(0)}_{n+1} \bigr) = 0;
$$

\item update the plastic deformation for the final stress state on the yield surface
$$ 
\bepsilon^{p~(1)}_{n+1} = \bepsilon^{p}_{n} + \Delta \bepsilon - \mathbb{E}^{-1} \bigl[ \bsigma^{(1)}_{n+1} - \bsigma_{n} \bigr];
$$

\item EXIT.

\end{enumerate}

\end{minipage}
}
\end{center}

\vspace*{3mm}

There are two \lq find' in the procedure explained in Box 1: the first is at point (\ref{puta}) and the second is at point (\ref{zorra}). 
Both correspond to a root-finding procedure for a scalar function (the yield function) of tensorial variable (the stress), which can be pursued with 
different numerical techniques, so that we have employed a bisection method. Regarding the \lq find' at point (\ref{puta}), the zero of $F$ is sought 
along the segment joining $\bsigma_{n}$ with $\bsigma^{\textup{trial}}_{n+1}$, while no directions are \textit{a-priori} prescribed for returning on the 
yield surface from the stress state $\bsigma^{(0)}_{n+1}$ at point (\ref{ueppla}). We propose to find the zero of 
$F \bigl(\bsigma^{(1)}_{n+1}, \bepsilon^{p~(0)}_{n+1} \bigr) = 0$ along the segment drawn from $\bsigma^{(0)}_{n+1}$ to the centre of mass of the 
yield surface, $\bsigma_{G}$ [defined by parameter $p_{G}$, eq. \eqref{baricentro}].

Note finally that the presented numerical algorithm has the inconvenient typical of explicit methods, for which there is a small discrepancy at the end of 
the procedure, in the sense that the stress point lies on a yield surface which does not correspond to the updated values of hardening. A procedure 
alternative to the centre-of-mass algorithm is introduced in the next section.

\section{The \lq cutoff-substepping' integration algorithm}
\label{cutoffsub}

As an alternative to the forward Euler procedure with \lq centre of mass' return correction introduced in the previous section, we propose an implicit 
integration scheme. Since the standard return mapping algorithm does not work in a zone of the 
stress-space, this zone can be delimited by introducing a cutoff plane orthogonal to the hydrostatic axis, so that a new algorithm can be set up in which 
the return mapping scheme is augmented of a substepping when the trial elastic stress falls within that zone. In particular, if the trial elastic solution 
$\bsigma^{\textup{trial}}$ falls on the same side of the plane as the starting point, the return mapping algorithm correctly converges (as demonstrated 
in Section \ref{come}), while, if it falls beyond the cutoff plane, an iterative subincrementation is performed, in which the strain increment $\Delta \bepsilon$ is  
subdivided and the return mapping is iteratively applied with successive updates of the BP yield function, so that, eventually, the entire initial step 
will be performed remaining within the correct stress zone.

\begin{figure}[!htb]
\centering
\includegraphics[width=0.60\columnwidth,keepaspectratio]{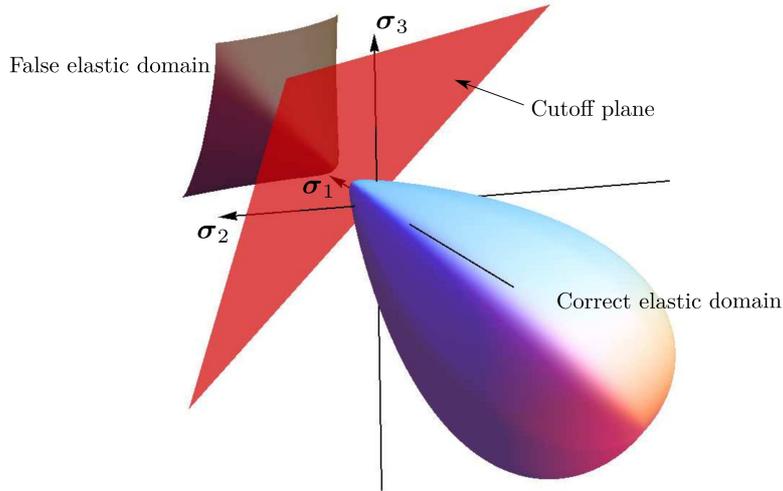}
\caption{Cutoff plane for the BP yield surface. Stress points where the return mapping algorithm works correctly are on the side of the plane where the 
yield surface lies. The false elastic domain is shown brown.}
\label{cutoff}
\end{figure}

The position of this cutoff plane depends on shape and size of the BP yield surface, see Fig. \ref{cutoff}, and can be determined as follows. 
%
%

\subsection{The squared BP yield function and the cutoff plane}
\label{come}

The squared BP yield function is obtained by squaring the terms in equation \eqref{effedip}, so that its meridian part (divided by $ p_{c} $) can be written 
as
\begin{equation}
\label{fpquadro}
\tilde{f}(\Phi) = M^{2} \left( \Phi-\Phi^{m} \right) \left[ 2(1-\alpha)\Phi + \alpha \right].
\end{equation}
The first and second derivatives of this function with respect to $\Phi$ are
\begin{equation}
\label{fpquadrod1}
\frac{ \diff \tilde{f}(\Phi) }{\diff \Phi} = M^{2} \left\{ 2(1-\alpha) \left[ 2\Phi - (1+m)\Phi^{m} \right] + \alpha \bigl( 1 - m \Phi^{m-1} \bigr) \right\},
\end{equation}
and
\begin{equation}
\label{fpquadrod2}
\frac{ \diff^{2} \tilde{f}(\Phi) }{\diff \Phi^{2}} = M^{2} \left\{ 2(1-\alpha) \left[ 2 - m(1+m) \Phi^{m-1} \right] - \alpha m(m-1) \Phi^{m-2} \right\},
\end{equation}
respectively. Note that the squared BP yield function is differentiable (its first and second derivatives are defined everywhere), but, in general, is no 
longer convex and displays a so-called \lq false elastic domain' (a nomenclature introduced by Brannon and Leelavanichkul \cite{bran}), visible in Fig.\ \ref{cutoff}. 
For this reason, the Newton-Raphson algorithm
\begin{equation}
\label{newton-raphson}
\Phi_{n+1} = \Phi_{n} - \dfrac{ \tilde{f}(\Phi_{n}) }{ \frac{ \diff \tilde{f}(\Phi) }{\diff \Phi} \Bigr|_{\Phi_{n}} } ,
\end{equation}
in general fails to converge. Nevertheless, it is possible to demonstrate that, for the squared BP yield function, a non-convex region exists in which the 
Newton-Raphson method still converges, despite the non-convexity. The region is delimited by the above-introduced cutoff plane, which position can be 
determined as follows.

\paragraph{Position of the cutoff plane}\

\begin{figure}[!htcb]
\centering
\includegraphics[width=90mm]{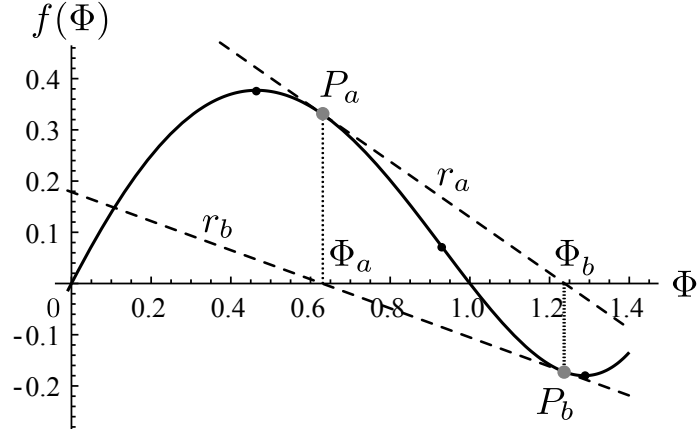}
\caption{Determination of the position of the cutoff plane. Local stationary (maximum and minimum) and inflection points are denoted by black spots, while 
the bounds of the non-convex region (in which the Newton-Raphson algorithm can be still used) are shown gray. The dashed lines $ r_{a} $ and $ r_{b} $ are 
the tangent lines to the meridian function at  the points $ P_{a} $ and $ P_{b} $, respectively. The graph $ \tilde{f}(\Phi) $ has been obtained with the 
following set of parameters: $ M = 1 $, $ m = 3 $, $ \alpha = 1.5 $, $ p_{c} = 100\text{ MPa} $, and $ c = 10\text{ MPa} $.}
\label{Intervallo-Convergenza-Newton-Raphson}
\end{figure}

Let us consider the situation sketched in Fig.\ \ref{Intervallo-Convergenza-Newton-Raphson}.
%
%
The generic points $P_{a} = \bigl( \Phi_{a}, \tilde{f}(\Phi_{a}) \bigr)$ and $P_{b} = \bigl( \Phi_{b}, \tilde{f}(\Phi_{b}) \bigr)$ lie on the meridian 
function, so that it is possible to calculate in those points the tangents
\begin{equation}
\label{fasci-rette-a}
r_{a} \colon \tilde{f}(\Phi) = \tilde{f}(\Phi_{a}) + \tilde{f}'(\Phi_{a}) (\Phi-\Phi_{a}),
\end{equation}
and 
\begin{equation}
\label{fasci-rette-b}
r_{b} \colon \tilde{f}(\Phi) = \tilde{f}(\Phi_{b}) + \tilde{f}'(\Phi_{b}) (\Phi-\Phi_{b}),
\end{equation}
where a prime denotes the derivative with respect to $\Phi$. If we impose that $\bigl( \Phi_{a}, 0 \bigr) \in r_{b}$ and $\bigl( \Phi_{b}, 0 \bigr) \in r_{a}$, 
we obtain the following non-linear algebraic system
\begin{equation}
\label{sist-fasci}
\begin{sistema}
\begin{aligned}
\tilde{f}(\Phi_{a}) + \tilde{f}'(\Phi_{a}) (\Phi_b-\Phi_{a}) &= 0, \\
\tilde{f}(\Phi_{b}) + \tilde{f}'(\Phi_{b}) (\Phi_a-\Phi_{b}) &= 0,
\end{aligned}
\end{sistema}
\end{equation}
with the unknowns $\Phi_{a}$ and $\Phi_{b}$; these values, that can be calculated numerically, define the region $[\Phi_{a}, \Phi_{b}]$ in which the 
Newton-Raphson algorithm can be still used, even though the squared BP yield function is not convex. 

As a conclusion, $\Phi_{b}$ is the value defining the position of the cutoff plane, to be used in the subincrementation scheme, as shown in Box 2 (note that  
$\Phi_{a}$ is not needed, since in the integration algorithm the trial elastic stress always lies outside the elastic domain). 

\begin{center}
\framebox{
\begin{minipage}{0.8\columnwidth}

\paragraph{\hfill Box 2: The \lq cutoff-substepping' integration algorithm \hfill}

\begin{enumerate}[(1)]

\item Given an initial state at step $n$, described by the variables  $\bsigma_{n}$, $\bepsilon^{e}_{n}$, $\bepsilon^{p}_{n}$ and given a strain 
increment $\Delta \bepsilon$;

\item Set $\Delta \bepsilon_i=\Delta \bepsilon$ and $m=1$ (where $m$ defines the substep interval); 

\item INITIALIZE: all variables are set equal to the value at the initial step $n$;

\item DO $i=1$, $m$; 

\item Evaluate the elastic trial solution 
$$ \bsigma^{\textup{trial}}_{n+1,i} = \bsigma_{n} + \mathbb{E} [\Delta \bepsilon_i] ;$$

\item Calculate $\Phi^{\textup{trial}}_{n+1,i} = \frac{p_{n+1,i}+c_n}{p_{c,n} + c_n}$ and $\Phi_b$ by solving eq. (\ref{sist-fasci});

\item Check position with respect to the cutoff plane

IF  $\Phi^{\textup{trial}}_{n+1,i} \leq \Phi_b$ GOTO Standard Return Mapping;

\item  Substepping procedure 

ELSE $m= 2 m$ AND $\Delta \bepsilon_i=\frac{\Delta \bepsilon}{m}$;

\item GOTO (3)

\end{enumerate}

\end{minipage}
}
\end{center}

\vspace*{11pt}

\section{The numerical performance: finite step accuracy}

The numerical performance of the centre-of-mass integration technique has been tested by comparing results obtained for a prescribed finite step of 
deformation (taken in different directions in the hyperspace of symmetric tensors as elucidated in Table \ref{tab01a}) with those obtained with the 
cutting-plane return-mapping technique \cite{sihug} applied to the \lq squared-version' of the BP yield surface, without subincrementation. 
In this way, it will become evident that for certain values of the trial elastic stress convergence will not occur for the latter algorithm. 
%
\begin{table}[!htcb]
\small
\newcommand{\ra}[1]{\renewcommand{\arraystretch}{#1}}
\centering
\ra{1.2}
\begin{tabular}{@{}llcc@{}}
\toprule
       &                                & \multicolumn{2}{c}{Deformation}\\
\cmidrule(l){3-4}
       &                                & $\Delta \varepsilon_{1}$ & $\Delta \varepsilon_{2} = \Delta \varepsilon_{3}$  \\
\midrule
Test 1 & Isotropic compression          & $-0.024$                 & $-0.024$                 \\
Test 2 & Isotropic traction             & $\phantom{-}0.00013714$  & $\phantom{-}0.00013714$  \\
Test 3 & Negative uniaxial deformation  & $-0.0080728$             & $0$                      \\
Test 4 & Positive uniaxial deformation  & $\phantom{-}0.00037312$  & $0$                      \\
Test 5 & Triaxial compression           & $-0.0092839$             & $-0.0185678$             \\
Test 6 & Triaxial extension             & $-0.006091$              & $-0.012182$              \\
\midrule
Test 7 & Shear                          & $\Delta \varepsilon_{1} = -\Delta \varepsilon_{2}$ & $\Delta \varepsilon_{3}$ \\
       &                                & $\phantom{-}0.00078408$                        & 0                        \\
\bottomrule
\end{tabular}
\caption{Deformation steps $\Delta \varepsilon$ used for comparing the performance of the centre-of-mass integration algorithm with the return mapping, the latter performed on the squared version of the BP yield function.}
\label{tab01a}
\end{table}

The comparison between the two integration algorithms has been performed by assuming:

\begin{itemize}

\item a form of the yield surface, namely,
$$
M = 0.26, \quad m = 2, \quad  \alpha = 1.99, \quad \beta = 0.12, \quad \gamma = 0.98, \quad p_c= 350 \text{ MPa}, \quad c= 2 \text{ MPa},
$$

\item elastic parameters in terms of Lam\'{e} constants 
$$
\lambda = 2669.49 \text{ MPa},  \qquad\mu = 4745.76 \text{ MPa}, 
$$

\item linear strain-hardening.

\end{itemize}

Note that the above parameters have been selected to be representative of a concrete-like material and the linear-hardening elastoplastic model has been 
implemented as a Umat routine for Abaqus (Ver. 6.10). 

\begin{figure}[!htcb]
\centering
\includegraphics[width=140mm]{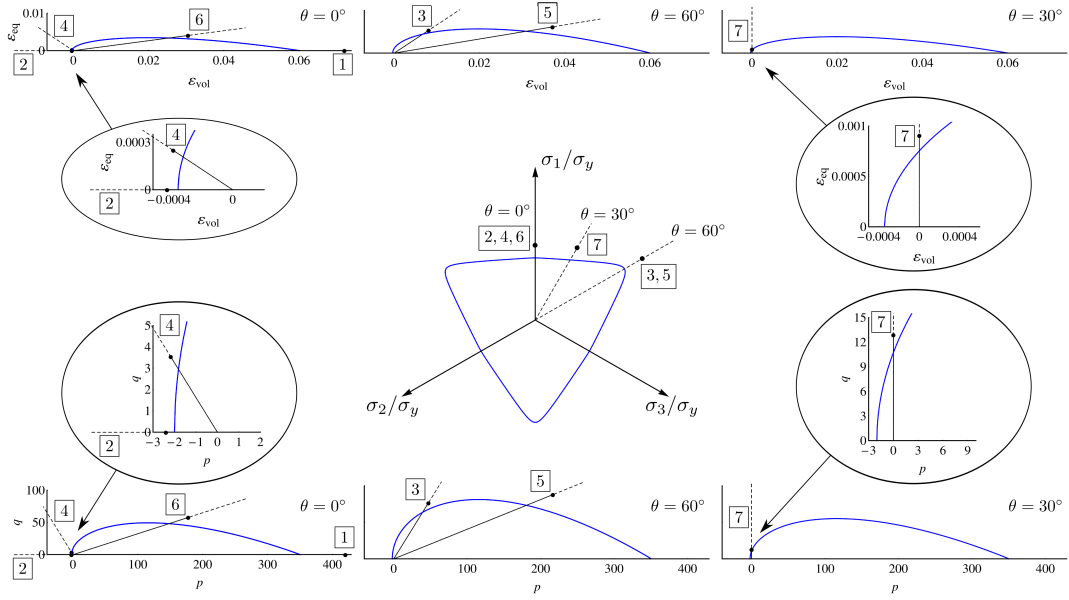}
\caption{Prescribed finite strain steps in the strain-space (upper part) and corresponding elastic trial stresses in the deviatoric plane (central part) and 
meridian plane (lower part) of the stress-space, for tests 1 to 7 reported in Table \ref{tab01a}. Finite steps are prescribed in such a way that the norm 
of the trial stress exceeds by 20\% the norm of the corresponding yield stress along the radial path from the origin to the trial stress.}
\label{path}
\end{figure}

The strain steps prescribed in Table \ref{tab01a} for testing the capability of the integration algorithms and the corresponding trial elastic stresses 
are reported together with the strain-space and stress-space representations of the BP meridian sections, respectively in the upper and lower parts of 
Fig.\ \ref{path}, where $\theta$ is the Lode's angle, eq. (\ref{lodazzo})$_3$. The trial elastic stresses in the deviatoric plane of the BP yield surface are reported in the central 
part of Fig.\ \ref{path}.
%

Note that the prescribed trial stresses have been given so that, in all cases, exactly the 20\% of its norm lies outside the elastic domain, 
$\lVert \bsigma^\text{trial} \lVert = 1.2 \times \lVert \bsigma_y \lVert$. 
Results, in terms of stress and plastic strain reached at the end of the procedure, are reported in Tab. \ref{tab01} for tests 1 to 6, while results of 
the test 7 are reported in Tabs. \ref{tab02} and \ref{tab02b}. In addition to the two algorithms under testing, a so-called \lq exact' result has also 
been included. This is obtained through successive subdivision of the strain increment into a sufficiently large number of subincrements to achieve 
convergence within a high tolerance (so that the relative error between the last two subincrements lies below $10^{-6}$). 

For the isotropic compression deformation path (\lq test  1') the return mapping algorithm fails to converge, as a consequence of the lack of convexity of the squared-version of the BP yield function, and therefore results are not reported in the table. 
%
\begin{table}[!htcb]
\small
\newcommand{\ra}[1]{\renewcommand{\arraystretch}{#1}}
\centering
\ra{1.2}
\begin{tabular}{@{}llccrcccr@{}}
\toprule
       &                  & \multicolumn{2}{c}{Stress}                                                    & Error 
       &                  & \multicolumn{2}{c}{Plastic strain}                                            & Error \\ 
\cmidrule(lr){3-5} \cmidrule(l){7-9}
       &                  & $\sigma_{1}$                      & $\sigma_{2} = \sigma_{3}$                 & \%
       &                  & $\varepsilon^{p}_{1}$             & $\varepsilon^{p}_{2}=\varepsilon^{p}_{3}$ & \%    \\
\midrule
Test 1 & Centre of mass   & $-384.8$                          & $-384.8$                                  & 0.05
       &                  & $-2.0103 \cdot 10^{-3}$           & $-2.0103 \cdot 10^{-3}$                   & 0.53  \\
       & Return mapping          & $-$                               & $-$                                       & $-$  
       &                  & $-$                               & $-$                                       & $-$   \\
       & Exact            & $-384.8$                          & $-384.8$                                  &
       &                  & $-2.0210 \cdot 10^{-3}$           & $-2.0210 \cdot 10^{-3}$                   &       \\
Test 2 & Centre of mass   & $\phantom{-}2.002$                & $\phantom{-}2.002$                        & 0.00
       &                  & $\phantom{-}2.2726\cdot 10^{-5}$  & $\phantom{-}2.2726\cdot 10^{-5}$          & 0.00  \\
       & Return mapping          & $\phantom{-}2.002$                & $\phantom{-}2.002$                        & 0.00
       &                  & $\phantom{-}2.2726\cdot 10^{-5}$  & $\phantom{-}2.2726\cdot 10^{-5}$          & 0.00  \\
       & Exact            & $\phantom{-}2.002$                & $\phantom{-}2.002$                        &  
       &                  & $\phantom{-}2.2726\cdot 10^{-5}$  & $\phantom{-}2.2726\cdot 10^{-5}$          &       \\
Test 3 & Centre of mass   & $-95.56$                          & $-25.88$                                  & 0.89
       &                  & $-3.7303\cdot 10^{-4}$            & $\phantom{-}3.5937\cdot 10^{-4}$          & 8.56  \\
       & Return mapping          & $-94.56$                          & $-25.23$                                  & 0.45   
       &                  & $-4.4081\cdot 10^{-4}$            & $\phantom{-}3.2748\cdot 10^{-4}$          & 4.25  \\
       & Exact            & $-94.89$                          & $-25.45$                                  &
       &                  & $-4.1833 \cdot 10^{-4}$           & $\phantom{-}3.3810\cdot 10^{-4}$          &       \\
Test 4 & Centre of mass   & $\phantom{-}4.029$                & $\phantom{-}0.616$                        & 0.23 
       &                  & $\phantom{-}3.3229 \cdot 10^{-5}$ & $\phantom{-}1.9628\cdot 10^{-5}$          & 2.14  \\
       & Return mapping          & $\phantom{-}4.015$                & $\phantom{-}0.628$                        & 0.31   
       &                  & $\phantom{-}3.4796\cdot 10^{-5}$  & $\phantom{-}1.8526\cdot 10^{-5}$          & 2.93  \\
       & Exact            & $\phantom{-}4.023$                & $\phantom{-}0.621$                        &
       &                  & $\phantom{-}3.3895 \cdot 10^{-5}$ & $\phantom{-}1.9159 \cdot 10^{-5}$         &       \\
Test 5 & Centre of mass   & $-191.4$                          & $-261.2$                                  & 0.04  
       &                  & $\phantom{-}5.2867 \cdot 10^{-4}$ & $-1.3881 \cdot 10^{-3}$                   & 0.16  \\
       & Return mapping          & $-191.5$                          & $-261.4$                                  & 0.09   
       &                  & $\phantom{-}5.3819\cdot 10^{-4}$  & $\phantom{-}1.0881\cdot 10^{-3}$          & 1.08  \\
       & Exact            & $-191.3$                          & $-261.2$                                  &
       &                  & $\phantom{-}5.2705 \cdot 10^{-4}$ & $-1.3947 \cdot 10^{-3}$                   &       \\
Test 6 & Centre of mass   & $-193.0$                          & $-146.1$                                  & 0.11  
       &                  & $-4.0150 \cdot 10^{-4}$           & $\phantom{-}7.4916\cdot 10^{-4}$          & 1.52  \\
       & Return mapping          & $-192.8$                          & $-145.9$                                  & 0.00  
       &                  & $-4.1020 \cdot 10^{-4}$           & $\phantom{-}7.3847\cdot 10^{-4}$          & 0.22  \\
       & Exact            & $-192.8$                          & $-145.9$                                  &
       &                  & $-4.1020 \cdot 10^{-4}$           & $\phantom{-}7.3847\cdot 10^{-4}$          &       \\
\bottomrule
\end{tabular}
\caption{Stress and plastic strain at the end of the finite step calculated with different algorithms for the strain and stress paths 1-6 of Table 
\ref{tab01a}, graphically represented in Fig. \ref{path}.}
\label{tab01}
\end{table}

\begin{table}[!htcb]
\small
\newcommand{\ra}[1]{\renewcommand{\arraystretch}{#1}}
\centering
\ra{1.2}
\begin{tabular}{@{}llcccr@{}}
\toprule
       &                & \multicolumn{3}{c}{Stress}                                   & Error \\
\cmidrule(l){3-6}
       &                & $\sigma_{1}$       & $\sigma_{2}$       & $\sigma_{3}$       & \%    \\
\midrule
Test 7 & Centre of mass & $\phantom{-}6.460$ & $-7.826$           & $-0.433$           & 0.61  \\
       & Return mapping        & $\phantom{-}6.441$ & $-7.741$           & $-0.354$           & 0.54  \\
       & Exact          & $\phantom{-}6.450$ & $-7.782$           & $-0.391$           &       \\
\bottomrule
\end{tabular}
\caption{Stress at the end of the finite step calculated with different algorithms for the strain and stress path 7 of Table \ref{tab01a}, graphically 
represented in Fig.\ \ref{path}.}
\label{tab02}
\end{table}

\begin{table}[!htcb]
\small
\newcommand{\ra}[1]{\renewcommand{\arraystretch}{#1}}
\centering
\ra{1.2}
\begin{tabular}{@{}llcccr@{}}
\toprule
       &                &  \multicolumn{3}{c}{Plastic strain}                                                                       & Error \\
\cmidrule(l){3-6}
       &                & $\varepsilon^{p}_{1}$             & $\varepsilon^{p}_{2}$             & $\varepsilon^{p}_{3}$             & \%    \\
\midrule
Test 7 & Centre of mass & $\phantom{-}7.4599\cdot 10^{-5}$  & $\phantom{-}1.1561\cdot 10^{-5}$  & $\phantom{-}1.6671\cdot 10^{-5}$  & 6.64  \\
       & Return mapping        & $\phantom{-}7.8884 \cdot 10^{-5}$ & $\phantom{-}4.9236 \cdot 10^{-4}$ & $\phantom{-}1.0745\cdot 10^{-5}$  & 5.92  \\
       & Exact          & $\phantom{-}7.6856 \cdot 10^{-5}$ & $\phantom{-}8.0822\cdot 10^{-6}$  & $\phantom{-}1.3524 \cdot 10^{-5}$ &   \\
\bottomrule
\end{tabular}
\caption{Plastic strain at the end of the finite step calculated with different algorithms for the strain and stress path 7 of Table \ref{tab01a}, 
graphically represented in Fig.\ \ref{path}.}
\label{tab02b}
\end{table}

\begin{figure}[!htb]
\centering
\includegraphics[width=100mm]{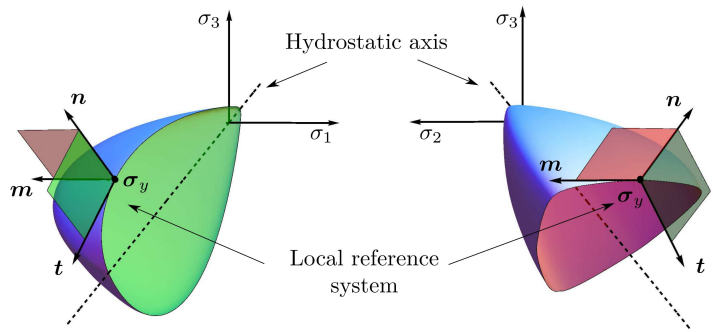}
\caption{Sections of the yield surface and local reference system employed for the construction of the iso-error maps.}
\label{planes}
\end{figure}

Iso-error maps have been plotted to display the error trend of the two algorithms in the stress-space for a set of different strain increments, chosen with 
the condition that the trial elastic solutions $\bsigma^{\textup{trial}}$ lie respectively in the meridian (denoted as $\boldsymbol{t}-\bn$ in Fig. \ref{planes} 
on the left) and deviatoric (denoted as $\boldsymbol{m}-\bn$ in Fig. \ref{planes} on the right) planes.

The iso-error maps plotting ranges have been chosen as follows: 
\begin{equation}
0 \leq \frac{\Delta\sigma^{\textup{trial}}_n}{|\bsigma_{y}|} \leq 0.2\ ,\quad 
-0.2\,\leq\,\frac{\Delta\sigma^{\textup{trial}}_t}{|\bsigma_{y}|} \leq 0.2 ,\quad  
-0.2\,\leq\,\frac{\Delta\sigma^{\textup{trial}}_m}{|\bsigma_{y}|} \leq 0.2.
\end{equation}
where $\bsigma_{y}$ is the considered stress at yielding 
\begin{equation}
\Delta \bsigma^{\textup{trial}} = \Delta\sigma^{\textup{trial}}_t \bt + \Delta\sigma^{\textup{trial}}_n \bn + \Delta\sigma^{\textup{trial}}_m \bm.
\end{equation}
%

\newpage

\begin{figure}[!htcb]
\centering
\includegraphics[width=118mm]{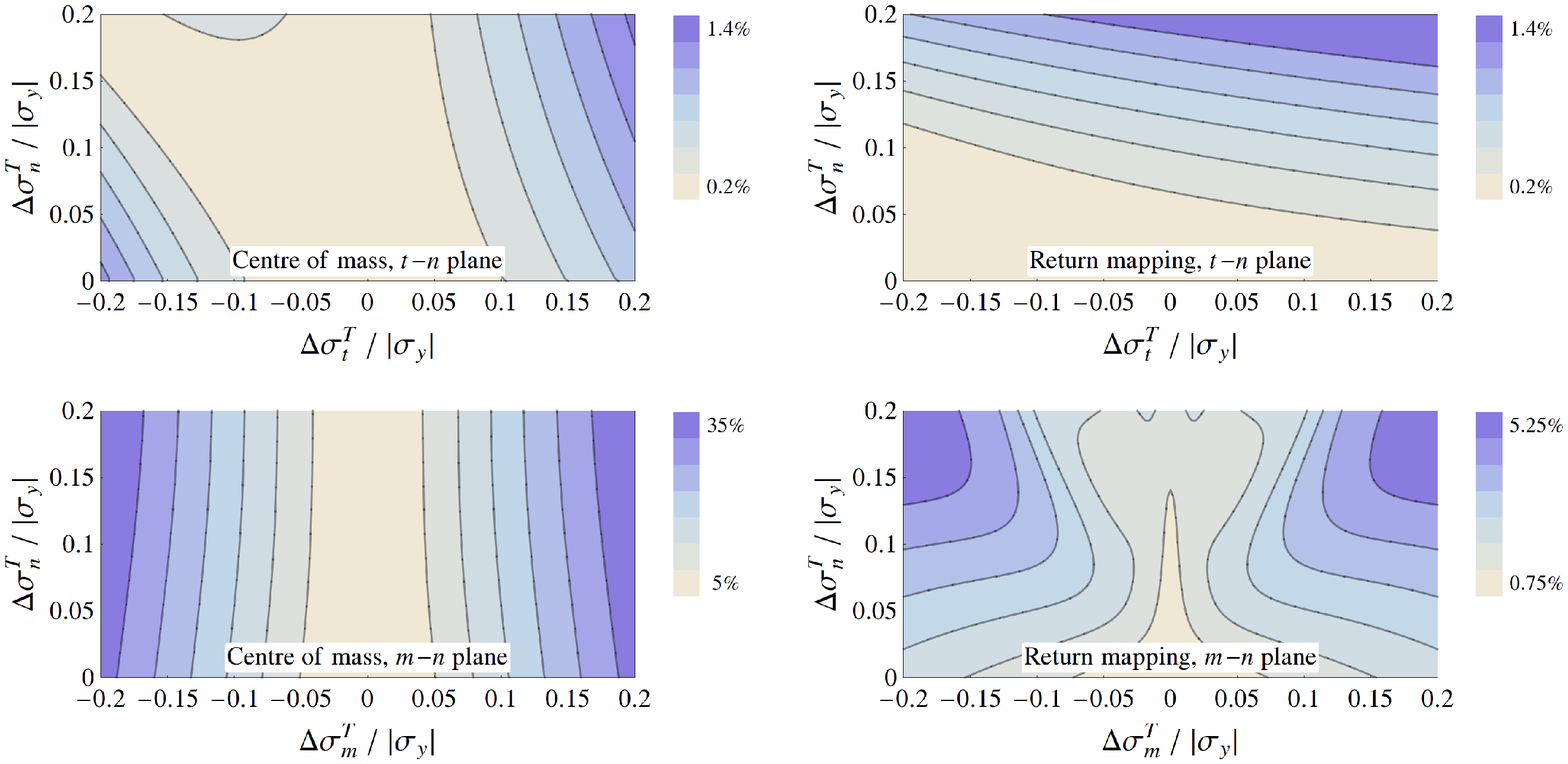}
\caption{Iso-error maps for Test 3 (see Tab. \ref{tab01a} and Fig. \ref{path}).}
\label{p3}
\end{figure}

\begin{figure}[!htcb]
\centering
\includegraphics[width=118mm]{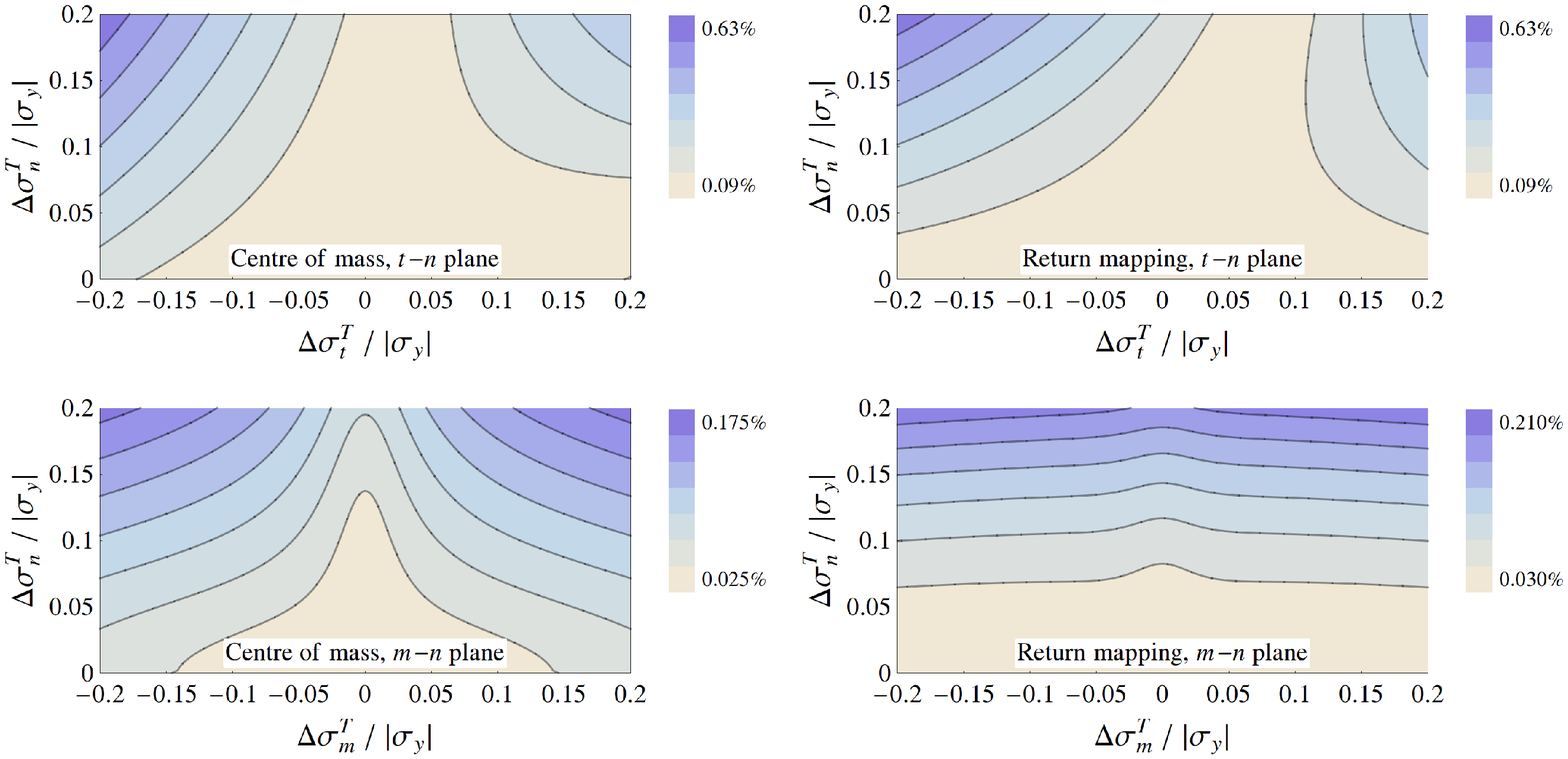}
\caption{Iso-error maps for Test 4 (see Tab. \ref{tab01a} and Fig. \ref{path}).}
\label{p4}
\end{figure}

\begin{figure}[!htcb]
\centering
\includegraphics[width=118mm]{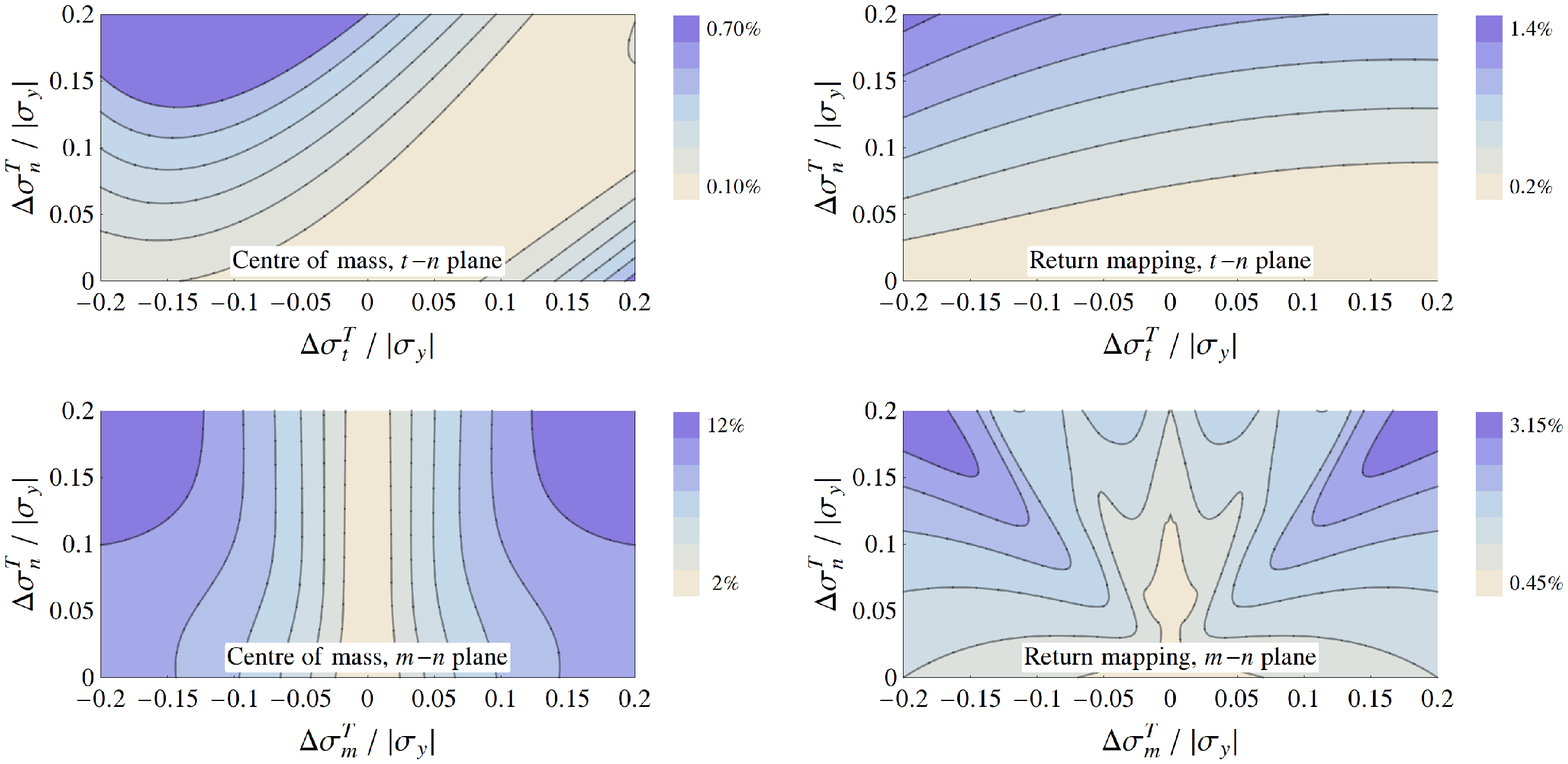}
\caption{Iso-error maps for Test 5 (see Tab. \ref{tab01a} and Fig. \ref{path}).}
\label{p5}
\end{figure}

\begin{figure}[!htcb]
\centering
\includegraphics[width=118mm]{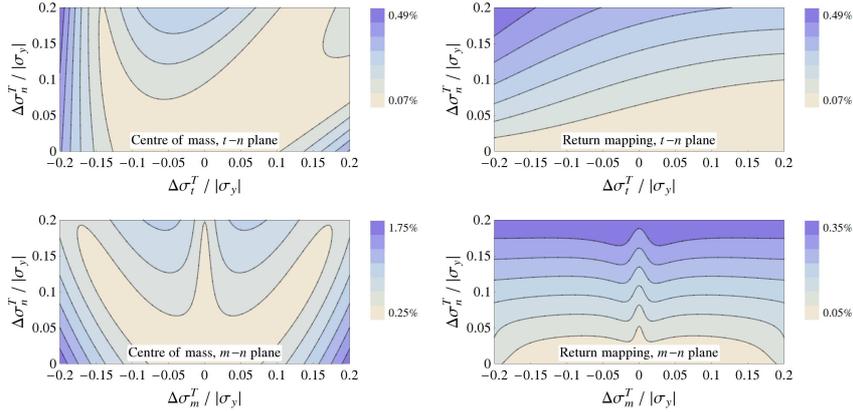}
\caption{Iso-error maps for Test 6 (see Tab. \ref{tab01a} and Fig. \ref{path}).}
\label{p6}
\end{figure}

The iso-error maps are reported in Fig.\ \ref{p3}--\ref{p6}, assuming as yield stresses $\bsigma_{y}$ those corresponding to the tests 3, 4, 5, and 6 of 
Tab.\ \ref{tab01a}, graphically represented in Fig.\ \ref{path}.  
%
%
%
%

It can be noted from Figs.\ \ref{p3} and \ref{p5} (bottom, left) that the centre-of-mass algorithm has a low accuracy when the yield stress $\bsigma_{y}$ 
lies near the corner of the deviatoric section (see Fig.\ \ref{path}, central part, tests 3 and 5) and the stress increment is not radial. On the other hand, 
the accuracy is high in both $\boldsymbol{t}-\bn$ and $\boldsymbol{m}-\bn$ planes, when the yield stress $\bsigma_{y}$ lies near the flat parts of this 
section (see Fig.\ \ref{path}, central part, tests 4 and 6), as shown in Figs.\ \ref{p4} and \ref{p6}.


\section{Comparison with semi-analytical solutions}

\begin{figure}[!htcb]
\centering
\includegraphics[width=140mm]{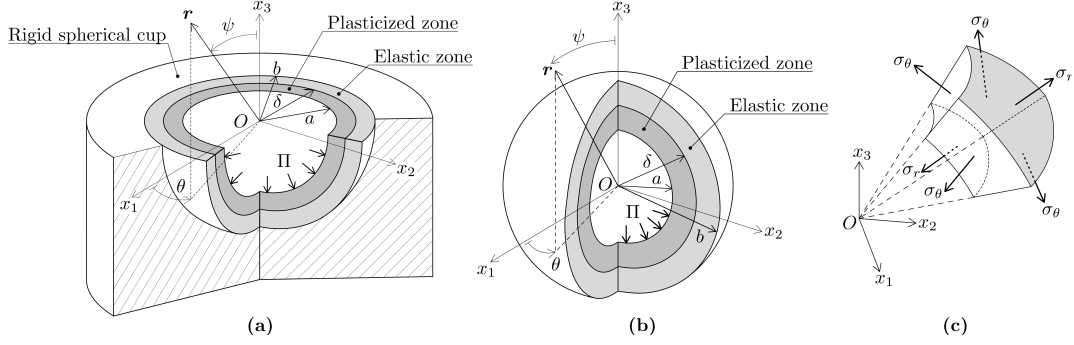}
\caption{Geometry for the compaction of a thick perfectly-plastic layer of ceramic powder againt a rigid cup (a) and for the expansion of a thick 
perfectly-plastic spherical shell under internal pressure (b). In both cases, the boundary of the plasticized zone is represented by $\delta$ which moves 
from $r=a$ to $r=b$ at increasing internal pressure $\Pi$. The reference system and stress components are shown in part (c).}
\label{sfera}
\end{figure}

Numerical results obtained by employing the proposed algorithms have been compared with semi-analytical solutions of a simple compaction problem and a 
deformation of a green body. In particular, in Section 
\ref{forming}, the forming of a thick perfectly-plastic layer of ceramic powder is considered, pressed against a rigid spherical cup, see Fig.\ \ref{sfera}a. Moreover, 
a thick spherical shell of a green body is considered in Section \ref{expansion}, subjected to an internal uniform pressure with a traction-free external 
boundary and expanded until collapse, corresponding to complete plasticization, see Fig.\ \ref{sfera}b.
Due to the spherical symmetry of both the problems, it is possible in both cases to obtain accurate semi-analytical solutions for the stress field by direct 
numerical integration of the equilibrium equations. 

These benchmark problems, differing only in the boundary conditions, are used to check the accuracy and efficiency of the proposed algorithms. They represent only model problem simulations of industrial processes and cannot be considered fully realistic, since hardening (and therefore the evolution of the yield surface) is neglected, so that the 
increase in cohesion is not taken into account.

The problem of the expansion of a thick spherical shell is interesting in itself, due to the applications in geotechnics, and it has been previously solved under a number 
of hypotheses \cite{hill,bila,cohen,rap,volok}, although never with the BP yield function. The problem of compaction of a layer of powder against a rigid cup was previously not addressed in analytically. 

For both problems, the inner and outer radii of the shell are denoted with $a$ and $b$ respectively, while the internal pressure is $\Pi$, which is assumed 
to increase from zero to the maximum value corresponding to the full plasticization of the shell. Since the geometry shows radial symmetry, 
we assume a spherical coordinate system. The solution is known in the case of perfect plasticity with the Tresca yield criterion \cite{hill}, so that 
our objective is to generalize the solution to the BP yield criterion. 

Due to the spherical symmetry, the stress and deformation depend only on the radius $r$. The non-vanishing deformation radial azimuthal, and polar components are respectively 
\begin{equation}
\label{eqgovern}
\varepsilon_{r} = \frac{\diff u}{\diff r} , \qquad \varepsilon_{\theta}=\varepsilon_{\phi} = \frac{u}{r} ,
\end{equation}
where $u$ is the radial displacement. The compatibility equation is
\begin{equation}
\label{compatibilità}
\varepsilon_{r} = \frac{\diff }{\diff r} (r\varepsilon_{\theta}) ,
\end{equation}
while the equilibrium equation in spherical coordinates is
\begin{equation}
\label{equilibrio}
\frac{\diff\sigma_{r} }{\diff r} + \frac{2}{r}(\sigma_{r}-\sigma_{\theta}) = 0 ,
\end{equation}
to be complemented by the boundary conditions. 

The elastic constitutive equations are
\begin{equation}
\label{cost}
\varepsilon_{r}      = E^{-1}(\sigma_r-2\nu\sigma_{\theta}) ,  \qquad
\varepsilon_{\theta} = E^{-1} \bigl[ (1-\nu)\sigma_{\theta}-\nu\sigma_{r} \bigr] ,
\end{equation}
where $E$ is the elastic Young modulus and $\nu$ the Poisson's ratio.
The Tresca yield criterion coincides (under the current assumptions) with the von Mises criterion, which can be written as
\begin{equation}
\label{costitutivo elastico2}
|\sigma_{r} - \sigma_{\theta}| - \sigma_{0} = 0 ,
\end{equation}
where $\sigma_{0}$ is the uniaxial yield stress, while the BP yield criterion \eqref{funzionazza} writes now in the following form
\begin{equation}
\label{snervamento BP}
F(\bsigma) = f \Bigl( \frac{\sigma_{r} + 2\sigma_{\theta}}{3} \Bigr) + \frac{ \lvert \sigma_{r}-\sigma_{\theta} \rvert} { g(\frac{\pi}{3})} = 0.
\end{equation}
\paragraph{The elastic solution.} Using equations \eqref{compatibilità}, \eqref{equilibrio} and \eqref{cost} we obtain
\begin{equation}
\label{ellapeppa}
\frac{1-\nu}{2} \frac{\diff}{\diff r} \left( \sigma_{r} + 2 \sigma_{\theta} \right) = 0 .
\end{equation}
This equation together with eq. (\ref{equilibrio}) forms a system of ODEs, which can be solved exactly and the solution is given by
\begin{equation}
\label{sol}
\sigma_r(r) = \frac{C_1}{3} + \frac{C_2}{r^3}, \qquad \sigma_\theta(r) = \frac{C_1}{3} - \frac{C_2}{2r^3},
\end{equation}
where $C_1$ and $C_2$ are constants to be defined through the boundary conditions. The associated deformation and displacement fields are obtained from 
(\ref{cost}) and (\ref{eqgovern}) and read
\begin{equation}
\varepsilon_r(r) = \frac{1}{E} \left[ (1 - 2\nu) \frac{C_1}{3} + (1 + \nu) \frac{C_2}{r^3} \right], \quad 
\varepsilon_\theta(r) = \frac{1}{E} \left[ (1 - 2\nu) \frac{C_1}{3} - (1 + \nu) \frac{C_2}{2r^3} \right],
\end{equation}
\begin{equation}
\label{uuu}
u(r) = \frac{1}{E} \left[ (1 - 2\nu) \frac{C_1}{3} r - (1 + \nu) \frac{C_2}{2r^2} \right],
\end{equation}

\subsection{Compaction of a thick layer of perfectly-plastic material obeying the BP yield condition against a rigid spherical cup}
\label{forming}

For the compaction problem of a thick layer against a rigid spherical cup,  Fig. \ref{sfera}a, the boundary conditions write as follows
\begin{equation}
\label{bordo}
\sigma_r \big|_{r=a} = -\Pi ,  \qquad   u \big|_{r=b} = 0,
\end{equation}
where $\Pi$ is the internal pressure. The material parameters defining the shape of the BP yield surface have been chosen to be representative of alumina 
powder (Piccolroaz et al., 2006), namely
$$
M = 1.1, \quad m = 2, \quad \alpha = 0.1, \quad \beta = 0.19, \quad \gamma = 0.9, \quad p_c= 40 \text{ MPa}, \quad c = 1.5 \text{ MPa}.
$$
Note that, since hardening and increasing of cohesion are neglected, we assume an initial state corresponding to an intermediate stage of a densification 
process.

\subsubsection{The elastic solution}

Initially the problem is purely elastic, which occurs when the internal pressure is sufficiently small, say, $\Pi \leq \Pi_{y}$, where $\Pi_{y}$ is defined as 
the inner pressure producing the initiation of yielding at the inner radius of the shell.

The solution (\ref{sol})--(\ref{uuu}) together with boundary conditions \eqref{bordo}, provides the following stress field within the thick spherical layer, 
$a \leq r \leq b$,
\begin{equation}
\label{elas1}
\sigma_r^\text{e}(\Pi,r) = 
-\frac{a^3 (1 + \nu) \Pi}{a^3 (1 + \nu) + 2b^3 (1 - 2\nu)} - \frac{2a^3b^3 (1 - 2\nu) \Pi}{a^3 (1 + \nu) + 2b^3 (1 - 2\nu)} \frac{1}{r^3}, 
\end{equation}
\begin{equation}
\label{elas2}
\sigma_\theta^\text{e}(\Pi,r) = 
-\frac{a^3 (1 + \nu) \Pi}{a^3 (1 + \nu) + 2b^3 (1 - 2\nu)} + \frac{a^3b^3 (1 - 2\nu) \Pi}{a^3 (1 + \nu) + 2b^3 (1 - 2\nu)} \frac{1}{r^3}.
\end{equation}
For the von Mises yield criterion, the critical yield pressure $\Pi_{y}$ is represented by the stress state satisfying
\begin{equation}
\lvert \sigma_{r}^\text{e} - \sigma_{\theta}^\text{e} \rvert = \sigma_{0},
\end{equation}
and can be evaluated as
\begin{equation}
\label{ellapeppa5}
\Pi_{y} = \frac{\sigma_{0}}{3} \left[ 2 + \frac{1 + \nu}{1 - 2\nu} \left(\frac{a}{b}\right)^{3} \right].
\end{equation}
In the following calculations $\nu = 0.26$ has been assumed. For the BP yield criterion, the critical yield pressure $\Pi_{y}$ corresponds to a 
stress state satisfying
\begin{equation}
\label{6.50}
\max_{a \leq r \leq b}\ F \bigl( \sigma_{r}^\text{e}(\Pi_{y},r), \sigma_{\theta}^\text{e}(\Pi_{y},r) \bigr) = 0,
\end{equation}
so that $\Pi_{y}$ can be evaluated as the numerical solution of the above equation and it can be numerically shown that the plasticization starts from the 
inner surface of the layer, $r = a$.

\subsubsection{The elasto-plastic solution}

The elasto-plastic solution holds for an internal pressure $\Pi > \Pi_{y}$, which implies both elastic and plastic deformation of the layer.
The plastic flow starts from the inner surface of the layer and propagates within a spherical region with inner radius $a$ and outer $\delta$ and moving toward $b$. The remaining part of the layer, namely, for $\delta \leq r \leq b$, behaves as an elastic layer with inner 
radius $\delta$ and outer $b$, subject to an internal pressure $\Pi_{\delta}$ at the interface with the plasticized zone.

Assuming that the yield pressure at the interface $r = \delta$ is $\Pi_{\delta}$, a generic yield criterion writes as
\begin{equation}
\label{ellapeppa7}
F \bigl( \sigma_{r}^\text{e}(\Pi_{\delta},\delta), \sigma_{\theta}^\text{e}(\Pi_{\delta},\delta) \bigr) = 0,
\end{equation}
which provides a relation between $\delta$ and $\Pi_{\delta}$. For example, the pressure at the interface for the von Mises criterion can be obtained from 
eq. \eqref{ellapeppa5} imposing $a = \delta$ as 
\begin{equation}
\label{ellapeppa6}
\Pi_{\delta} = \frac{\sigma_{0}}{3} \left[ 2 + \frac{1 + \nu}{1 - 2\nu} \left(\frac{\delta}{b}\right)^{3} \right].
\end{equation}
whereas for the BP criterion the pressure $\Pi_{\delta}$ has to be evaluated numerically.

The solution for the elastic zone ($\delta \leq r \leq b$) can be obtained from eqs. \eqref{elas1} and \eqref{elas2} where $a$ and $\Pi$ are replaced, 
respectively, by $\delta$ and $\Pi_{\delta}$ which are given by \eqref{ellapeppa7}, so that the stresses become
\begin{equation}
\label{pippo1}
\sigma_r^\text{ep}(r) = -\frac{\delta^3 (1 + \nu) \Pi_\delta}{\delta^3 (1 + \nu) + 2b^3 (1 - 2\nu)} - 
\frac{2\delta^3b^3 (1 - 2\nu) \Pi_\delta}{\delta^3 (1 + \nu) + 2b^3 (1 - 2\nu)} \frac{1}{r^3}, 
\end{equation}
\begin{equation}
\label{pippo2}
\sigma_\theta^\text{ep}(r) = -\frac{\delta^3 (1 + \nu) \Pi_\delta}{\delta^3 (1 + \nu) + 2b^3 (1 - 2\nu)} + 
\frac{\delta^3b^3 (1 - 2\nu) \Pi_\delta}{\delta^3 (1 + \nu) + 2b^3 (1 - 2\nu)} \frac{1}{r^3},
\end{equation}
Hence the elastic part of the solution is known as the relation between the radius $\delta$ and the pressure $\Pi_{\delta}$ is known.

The solution for the plasticized zone ($a \leq r \leq \delta$) is obtained from the algebraic-differential system composed by the equilibrium equations 
\eqref{equilibrio}, the boundary conditions \eqref{bordo}, and the yield condition \eqref{costitutivo elastico2} or \eqref{snervamento BP} (depending on 
the criterion assumed). This system writes as
\begin{equation}
\label{sistdiff}
\begin{sistema}
\ds \frac{\diff \sigma_{r}}{\diff r } + \frac{2}{r} \left( \sigma_{r}-\sigma_{\theta} \right) = 0, \\[4mm]
F \bigl( {\sigma}_{r} (r), {\sigma}_{\theta} (r) \bigr) = 0, \\[4mm]
\sigma_{r} \big|_{r=a} = -\Pi,  \\[4mm]
\sigma_{r} \big|_{r=\delta} = -\Pi_{\delta} ,
\end{sistema}
\end{equation}
which has been solved analytically for von Mises yield and numerically for the BP yield function. In particular, the system \eqref{sistdiff} admits for 
von Mises the following solution
\begin{equation}
\label{ellapeppafin}
\sigma_{r}^\text{ep}(r) = 
-\frac{\sigma_{0}}{3} \left[ 2 + \frac{1 + \nu}{1 - 2\nu} \left(\frac{\delta}{b}\right)^{3} + 6\log\left(\frac{\delta}{r}\right) \right], \quad 
\sigma_{\theta}^\text{ep}(r) = 
-\frac{\sigma_{0}}{3} \left[ -1 + \frac{1 + \nu}{1 - 2\nu} \left(\frac{\delta}{b}\right)^{3} + 6\log\left(\frac{\delta}{r}\right) \right],
\end{equation}
and the relation between $\delta$ and the internal pressure $\Pi$ writes as
\begin{equation}
\label{ellapeppa13}
\Pi = \frac{\sigma_{0}}{3} \left[ 2 + \frac{1 + \nu}{1 - 2\nu} \left(\frac{\delta}{b}\right)^{3} + 6\log\left(\frac{\delta}{a}\right) \right],
\end{equation}
which is a nonlinear relation. Once a fixed value of the radius $\delta$, representing the amplitude of the plasticized zone, is 
chosen, it is possible to obtain the internal pressure $\Pi$ and the stresses in every part of the layer, namely for $a \le r \le b$.

\begin{figure}[!htb]
\centering
\includegraphics[width=0.8\columnwidth,keepaspectratio]{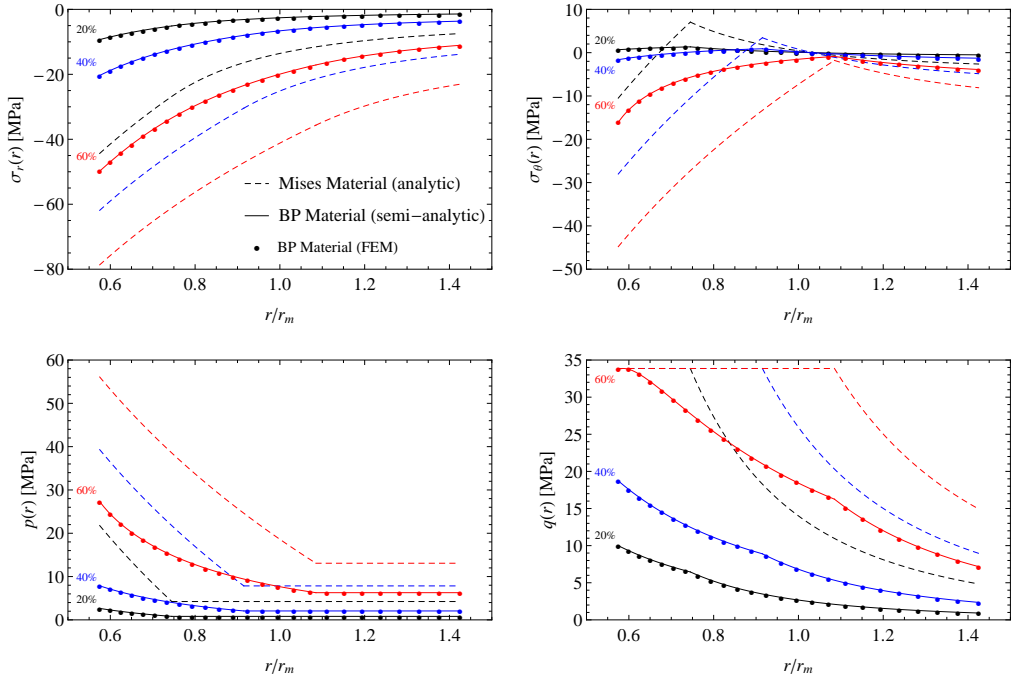}
\caption{Compaction of a perfectly-plastic thick layer, obeying von Mises and BP yield conditions, against a rigid spherical cup, representative of a 
ceramic powder. Upper part: radial (left) and polar (right) stress components as functions of the dimensionless radial position. 
Lower part: mean stress $p$ (left) and deviatoric invariant $q$ (right) as functions of the dimensionless radial position.
Note that for the von Mises criterion $\sigma_0 = 33.86$ MPa has been chosen, so that the von Mises cylinder is circumscribed around the BP surface in the 
stress space.}
\label{sferazza}
\end{figure}

Results in terms of radial and polar stress components and the two stress invariants $p$ and $q$ are reported in Fig.\ \ref{sferazza} as functions of the through-thickness radius (divided by the mean radius $r_m=(a+b)/2$ of the spherical layer), together 
with the numerical results obtained with the two proposed algorithms. Three different plastic boundaries $\delta$ have been considered (corresponding to 
the 20\%, 40\% and 60 \% of the thickness) for both von Mises and the BP yield criterion. 
Results presented in the figure fully support the validity of the proposed numerical algorithms, which have given coincident results, superimposed on 
the semi-analytical solution.


\subsection{The expansion of a perfectly plastic thick shell obeying the BP yield condition}
\label{expansion}

For the problem of expansion of a thick spherical shell subjected to an internal uniform pressure, Fig.\ \ref{sfera}b, the boundary conditions are as follows
\begin{equation}
\label{bordo2}
\sigma_r \big|_{r=a} = -\Pi ,  \qquad   \sigma_r \big|_{r=b} = 0 ,
\end{equation}
where $\Pi$ is the internal pressure and the outer boundary is assumed traction-free. The material parameters defining the shape of the BP yield surface have 
been chosen to be representative of a partially densified ceramic powder, namely
$$
M = 1.33, \quad m = 2, \quad \alpha = 1, \quad \beta = 1, \quad \gamma = 0, \quad p_c= 150 \text{ MPa}, \quad c = 150 \text{ MPa}.
$$
The solution of this problem can be obtained with the same method as that described in Sec. \ref{forming}, since only the boundary conditions are different.

The elastic solution, valid until the internal pressure is sufficiently small, $\Pi \leq \Pi_{y}$, is given by
\begin{equation}
\sigma_{r}^\text{e}(r) = \frac{\Pi}{ \left( \frac{b}{a} \right)^{3}-1 } \left[ 1-\left( \frac{b}{r} \right)^3 \right], \qquad
\sigma_{\theta}^\text{e}(r) = \frac{\Pi}{ \left( \frac{b}{a} \right)^{3}-1 } \left[ 1+\frac{1}{2} \left( \frac{b}{r} \right)^{3} \right].
\end{equation}
For the von Mises yield criterion, $\lvert \sigma_{r}^\text{e} - \sigma_{\theta}^\text{e} \rvert = \sigma_{0}$, the critical yield pressure $\Pi_{y}$ is 
obtained as
\begin{equation}
\label{ellapeppa5b}
\Pi_{y} = \frac{2}{3} \sigma_{0} \left[ 1 - \left(\frac{a}{b}\right)^{3} \right],
\end{equation}
whereas for the BP yield criterion, the critical yield pressure $\Pi_{y}$ is obtained by solving eq. (\ref{6.50}) and it can be numerically proven that 
the plasticization starts from the inner surface of the shell.

The elasto-plastic solution holds for an internal pressure $\Pi > \Pi_{y}$, which implies both elastic and plastic deformation of the shell.
The plastic flow starts from the inner surface of the shell and propagates within a spherical region with inner radius $a$ and outer $\delta$ and 
moving toward $b$. The remaining part of the shell, namely, for $\delta \leq r \leq b$, behaves as an elastic shell with inner radius 
$\delta$ and outer $b$, subject to an internal pressure $\Pi_{\delta}$ at the interface with the plasticized zone.

The relation between $\delta$ and $\Pi_{\delta}$ is obtained by solving eq. (\ref{ellapeppa7}). For the von Mises criterion $\Pi_\delta$ is obtained as
\begin{equation}
\label{ellapeppa6b}
\Pi_{\delta} = \frac{2}{3} \sigma_{0} \left[ 1 - \left( \frac{\delta}{b} \right)^{3} \, \right],
\end{equation}
whereas for the BP criterion the pressure $\Pi_{\delta}$ has to be evaluated numerically.

The solution for the elastic zone, $\delta \leq r \leq b$, is given by
\begin{equation}
\label{pippob}
\sigma_{r}^\text{ep}(r) = \frac{ \Pi_{\delta} }{ \left( \frac{b}{\delta} \right)^{3}-1 } \left[ 1-\left( \frac{b}{r} \right)^3, \right], \qquad
\sigma_{\theta}^\text{ep}(r) = \frac{ \Pi_{\delta} }{ \left( \frac{b}{\delta} \right)^{3}-1 } \left[ 1+\frac{1}{2} \left( \frac{b}{r} \right)^{3} \right].
\end{equation}

The solution for the plasticized zone, $a \leq r \leq \delta$, is obtained from the algebraic-differential system (\ref{sistdiff}). This system has a 
solution with closed form for the simple case of von Mises yield criterion; in this case the stresses take the form
\begin{equation}
\label{ellapeppafin2}
\sigma_{r} = -\frac{2}{3} \sigma_{y} \left[ 1-\left( \frac{\delta}{b} \right)^{3} + \ln \left( \frac{\delta}{r} \right)^{3} \right], \qquad
\sigma_{\theta} = \frac{1}{3} \sigma_{y} \left[ 1+2\left( \frac{\delta}{b} \right)^{3} - 2\ln \left( \frac{\delta}{r} \right)^{3} \right] ,
\end{equation}
and the relation between $\delta$ and the internal pressure $\Pi$ writes as
\begin{equation}
\label{ellapeppa13b}
\Pi = \frac{2}{3} \sigma_{y} \left[ 1-\left( \frac{\delta}{b} \right)^{3} + \ln \left( \frac{\delta}{a} \right)^{3} \right].
\end{equation}
Once a fixed value of the radius $\delta$ representing the amplitude of the plasticized zone is 
chosen, it is possible to obtain the internal pressure $\Pi$ and the stresses in every part of the shell, namely for $a \le r \le b$.

\begin{figure}[!htb]
\centering
\includegraphics[width=0.8\columnwidth,keepaspectratio]{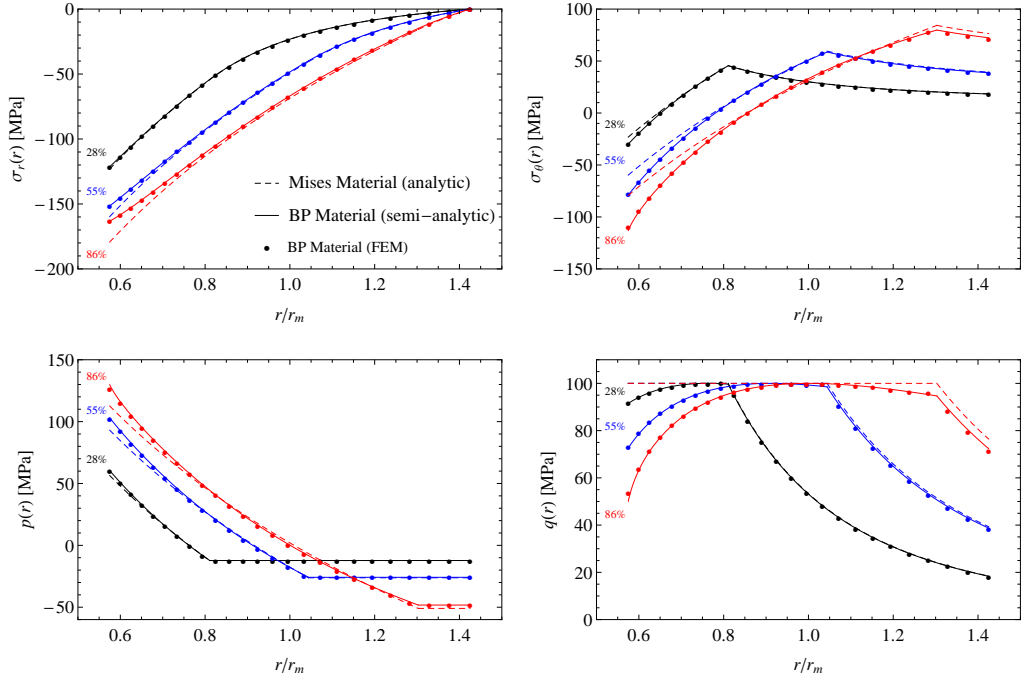}
\caption{Expansion of a perfectly-plastic thick spherical shell, obeying von Mises and BP yield conditions, representative of a green body. Upper part: 
radial (left) and polar (right) stress components as functions of the dimensionless radial position. Lower part: mean stress $p$ (left) and deviatoric invariant $q$ (right) 
as functions of the dimensionless radial position. Note that for the von Mises criterion $\sigma_0 = 100$ MPa has been chosen, so that the von Mises cylinder is circumscribed 
around the BP surface in the stress space.}
\label{sfera_3}
\end{figure}

Results in terms of radial and polar stress components and the two stress invariants $p$ and $q$ are reported in Fig.\ \ref{sfera_3} as functions of the through-thickness radius (divided by the mean radius $r_m=(a+b)/2$ of the thick shell), together 
with the numerical results obtained with the two proposed algorithms. Three different plastic boundaries $\delta$ have been considered (corresponding to 
the 28\%, 55\% and 86\% of the thickness) for both von Mises and the BP yield criterion. 
%
Again the two proposed algorithms have given coincident values, superimposed with the semi-analytical solution, thus confirming once more the validity of the presented numerical approaches.

\newpage

\section{Conclusions}

Two different algorithms have been presented for the integration of elastoplastic constitutive equations based on the BP yield function (useful in simulating forming of ceramic powders, deformation of granular bodies and, more in general, damage and failure of rock-like materials). 
One of the algorithms is based on a forward Euler scheme and the other on a return mapping technique with substepping. Though the former method is in some 
cases faster, the latter combines accuracy with efficiency and is eventually preferred.

\section*{Acknowledgments}
Part of this work was prepared during secondment of D.B. at Enginsoft S.p.A. (TN). Discussions with D. Blanco, G. Borzi, M. Gabrielli,
L. Gatti are gratefully acknowledged. 
A.P., D.B. and L.A. gratefully acknowledge financial support from European Union FP7-INTERCER2 project under contract number PIAP-GA-2011-286110.
M.P. gratefully acknowledges financial support from European Union FP7-HOTBRICKS project under contract number PIAPP-GA-2013-609758.


\begin{thebibliography}{99}

\setlength{\itemsep}{-1.5mm}

{\footnotesize

\bibitem{ang} Angelillo, M., Cardamone, L. and Fortunato, A. (2010) A numerical model for masonry-like structures. 
{\it J. Mech. Materials Structures} 5, 583-615.

\bibitem{ba} Babua, R.R., Benipal, G.S. and Singh, A.K. (2005) Constitutive modelling of concrete:  an overview. {\it Asian J. Civil Eng. (Building and Housing)} 6, 211-246.

\bibitem{bai} Bai, Y. and Wierzbicki, T. (2008) A new model of metal plasticity and fracture with pressure and Lode dependence. \IJP 24, 1071-1096.

\bibitem{bier}
Bier, W. and Hartmann, S. (2006)
A finite strain constitutive model for metal powder compaction using a unique and convex single surface yield function,
\EJMA 25, 1009-1030.

\bibitem{bigoni} Bigoni, D. (2012) {\it Nonlinear Solid Mechanics. Bifurcation theory and material instability}, Cambridge University Press.

\bibitem{bila} Bigoni, D. and Laudiero, F. (1989), The quasi-static finite cavity expansion in a non-standard elasto-plastic medium, Int. J. Mech. Sci., 31, 825-837.

\bibitem{bigp} Bigoni, D. and Piccolroaz, A. (2004) Yield criteria for quasibrittle and frictional materials.  \IJSS 41, 2855-2878.

\bibitem{bo} Bolchoun, A., Kolupaev, V.A., Altenbach, H. (2011) Convex and Non-convex Flow Surfaces. Forschung im Ingenieurwesen-Engineering Research 75, 73-92.

\bibitem{bosi} Bosi, F., A. Piccolroaz, M. Gei, F. Dal Corso, A. Cocquio and Bigoni, D. 2013. 
Experimental investigation of the elastoplastic response of aluminum silicate spray dried powder during cold compaction. {\it J. Europ. Ceramic Soc.} Submitted. 

\bibitem{bran} Brannon, R.M. and Leelavanichkul, S. (2010) Received: A multi-stage return algorithm for solving the classical damage component of constitutive models for rocks, ceramics, and other rock-like media. \IJF 163, 133–149.

\bibitem{cohen} Cohen T., Masri, R. Durban D. (2009) Analysis of circular hole expansion with generalized yield criteria, \IJSS 46, 3643-3650.  

\bibitem{conti} Conti, R., Tamagnini, C., DeSimone, A. (2013) Critical softening in Cam-Clay plasticity: Adaptive viscous regularization, dilated time and numerical integration across stress–strain jump discontinuities. \CMAME 258, 118–133. 

\bibitem{copp1}  Coppola, T., Cortese, L. and Folgarait, P. (2009) The effect of stress invariants on ductile fracture limit in steels. \EFM 76, 1288-1302.

\bibitem{copp} Coppola, T. and Folgarait, P. (2007) The influence of stress invariants on ductile fracture strain in steels. (in Italian) {\it Proc. XXXVI AIAS Congress}, Sept. 4-8, 2007.

\bibitem{dal1} Dal Maso, G., DeSimone, A. and Solombrino, F. (2012) Quasistatic evolution for Cam-Clay plasticity: properties of the viscosity solution. 
{\it Calculus Variat. Part. Diff. Equat.} 44, 495-541. 

\bibitem{dal} Dal Maso, G., Demyanov, A. and DeSimone, A. (2007) Quasistatic Evolution Problems for Pressure-sensitive Plastic Materials. {\it Milan J. Math.} 75, 117-134.

\bibitem{dalpino} Dal Pino, R., Narducci, P. and Royer-Carfagni, G. (1999) A SEM investigation on fatigue damage of marble. {\it J. Materials Sci. Letters} 18, 1619-1622.

\bibitem{defaveri} De Faveri, S., Freddi, L. and Paroni, R. (2013) No-tension bodies: A reinforcement problem. 
{\it Europ. J. Mech. A/Solids} 39, 163-169. 

\bibitem{des} Descamps, F. and Tshibangu, J.P. (2007) Modelling the Limiting Envelopes of Rocks in the Octahedral Plane. {\it Oil \& Gas Science and Technology - Rev. IFP}, 62, 683-694.

\bibitem{dr} DorMohammadi, H. and Khoei, A.R. (2008) A three-invariant cap model with isotropic-kinematic hardening rule and associated plasticity for granular materials. \IJSS 45, 631-656.

\bibitem{ebno} Ebnoether, F. and Mohr, D. (2013) Predicting ductile fracture of low carbon steel sheets: Stress-based versus mixed stress/strain-based Mohr-Coulomb model. \IJSS 50, 1055-1066.

\bibitem{ha} Hartmann, S. and Bier, W. (2008) High-order time integration applied to metal powder plasticity. \IJP  24, 17-54.

\bibitem{hei} Heisserer, U., Hartmann, S., D\"{u}ster, A., Bier, W., Yosibash, Z. and Rank, E. (2008) p-FEM for finite deformation powder compaction. \CMAME 197, 727-740.

\bibitem{hill}
Hill, R., (1950). 
The Mathematical Theory of Plasticity. Clarendon Press, Oxford.

\bibitem{hu} Hu, W. and Wang, Z.R. (2005) Multiple-factor dependence of the yielding behavior to isotropic ductile materials. {\it Comput. Mat. Sci.} 32, 31-46.

\bibitem{lex} Laydi, M.R and Lexcellent, C. (2010) Yield criteria for shape memory materials: convexity conditions and surface transport. \MMS 15, 165-208.

\bibitem{le} Lavernhe-Taillard, K., Calloch, S., Arbab-Chirani, S. and Lexcellent, C. (2009) Multiaxial Shape Memory Effect and Superelasticity. {\it Strain} 45, 77-84.

\bibitem{lex2002}
Lexcellent, C., Vivet, A., Bouvet, C., Calloch, S., Blanc, P. (2002)
Experimental and numerical determinations of the initial surface of phase transformation under biaxial loading in some polycrystalline shape-memory alloys,
\JMPS 50, 2717-2735.

\bibitem{mai}
Maiolino, S. (2005)
Proposition of a general yield function in geomechanics,
{\it Comptes Rendus Mecanique} 333, 279-284.

\bibitem{miro} Mirone, G. and Corallo, D. (2013) Stress-strain and ductile fracture characterization of an X100 anisotropic steel: Experiments and modelling. \EFM 102, 118-145.

\bibitem{mort}
Mortara, G. (2008)
A new yield and failure criterion for geomaterials,
{\it Geotechnique} 58, 125-132.

\bibitem{paluz} Paluszny, A. and Matthai, S.K. (2009) Numerical modeling of discrete multi-crack growth applied to pattern formation in geological brittle media. 46, 3383-3397. 

\bibitem{pibi} Piccolroaz, A. and Bigoni, D. (2009) Yield criteria for quasibrittle and frictional materials: a generalization to surfaces with corners. \IJSS 46, 3587-3596.

\bibitem{pibiga1}
Piccolroaz, A., Bigoni, D. and Gajo, A. (2006)
An elastoplastic framework for granular materials becoming cohesive through mechanical densification. Part. I - small strain formulation.
\EJMA 25, 334-357.

\bibitem{pibiga2}
Piccolroaz, A., Bigoni, D. and Gajo, A. (2006)
An elastoplastic framework for granular materials becoming cohesive through mechanical densification. Part. II - the formulation of elastoplastic coupling at large strain.
\EJMA 25, 358-369.

\bibitem{pie} Pietryga, M.P., Vladimirov, I.N., and Reese, S. (2012) A finite deformation model for evolving flow anisotropy with distortional hardening including experimental validation. \MOM 44, 163-173. 

\bibitem{podo1} Podg\'{o}rski, J. (1984) Limit state condition and the dissipation function for isotropic materials. {\it Arch. Mech. Soc.} 36, 323-342.

\bibitem{podo2} Podg\'{o}rski, J. (1985) General failure criterion for isotropic media. {\it J. Eng. Mech. ASCE} 111, 188-199.

\bibitem{rani98}
Raniecki, B. and Lexcellent, C. (1998)
Thermodynamics of isotropic pseudoelasticity in shape memory alloys,
\EJMA 17, 185-205.

\bibitem{rani} Raniecki, B. and Mr\'{o}z, Z. (2008) Yield or martensitic phase transformation conditions and dissipation functions for isotropic, pressure-insensitive alloys exhibiting SD effect. \AMEC 195, 81-102.

\bibitem{rap} Rapoport, L., Katzir, Z. and Rubin, M.B. (2011) Termination of the starting problem of dynamic expansion of a spherical cavity in an infinite elastic-perfectly-plastic medium. {\it Wave Motion} 48, 441-452.

\bibitem{saint} Saint-Sulpice, L., Arbab Chirani, S. and Calloch, S. (2009) A 3D super-elastic model for shape memory alloys taking into account progressive strain under cyclic loadings. \MOM 41, 12-26.

\bibitem{sed} Sedlak, P., Frost, M., Benesova, B., Ben Zineb, T. and Sittner, P. (2012) Thermomechanical model for NiTi-based shape memory alloys including R-phase and material anisotropy under multi-axial loadings. \IJP 39, 132-151. 

\bibitem{she} Sheldon, H.A., Barnicoat, A.C. and Ord, A. (2006) Numerical modelling of faulting and fluid flow in porous rocks: An approach based on critical state soil mechanics. {\it J. Struct. Geol.} 28, 1468-1482.

\bibitem{sihug} Simo, J. and Hughes, T.J.R. (1987) General return mapping algorithms for rate-independent plasticity. In: Desai, C.S. et al. (Eds.), Constitutive Laws for Engineering Materials: Theory and Applications. Elsevier Science Publ. Co., 221 –231.

\bibitem{stupk} Stupkiewicz, S., Piccolroaz, A.,  and Bigoni, D. (2013) Elastoplastic coupling to model cold ceramic powder compaction. {\it J. Europ. Ceramic Soc.} Submitted.

\bibitem{ta} Taillard, K., Arbab Chirani, S. Calloch, S. and Lexcellent, C. (2008) Equivalent transformation strain and its relation with martensite volume fraction for isotropic and anisotropic shape memory alloys. \MOM 40, 151-170.

\bibitem{volok} Volokh K.Y. (2011) Cavitation instability in rubber. {\it Int. J. Appl. Mech.} 3, 299-311. 

\bibitem{wang} Wang, S.Y., Sloan, S.W., Abbo, A.J., Masia, M.J., and Tang, C.A. (2012)  Numerical simulation of the failure process of unreinforced masonry walls due to concentrated static and dynamic loading. \IJSS 49, 377-394. 

\bibitem{wier} Wierzbicki, T., Bao, Y., Lee, Y-W., Bai, Y. (2005) Calibration and evaluation of seven fracture models. \IJMS 47, 719-743.

\bibitem{zho} Zhou, J., Yang, X., Yang, Z., Li, H., Zhou, H. (2013) Micromechanics damage modeling of brittle rock failure processes under compression. {\it Int. J. Comput. Meth.} 10, 1350034. 


}

\end{thebibliography}
\end{document}